\documentclass[a4paper,twocolumn,secnumarabic, amssymb, amsmath,nofootinbib, tightenlines,nobibnotes, showpacs, aps, prl]{revtex4-1}
\usepackage{booktabs}
\usepackage{graphicx}
\usepackage{subfigure}

\begin{document}


\title{The inverse scattering problem at fixed energy based on the Marchenko equation for an auxiliary Sturm-Liouville operator}

\author{Tam\'as P\'almai}
\email{palmai@phy.bme.hu}
\author{Barnab\'as Apagyi}
\email{apagyi@phy.bme.hu}
\affiliation{Department of Theoretical Physics, Budapest University of Technology and Economics, H-1111 Budapest, Hungary}


\begin{abstract}

A new approach is proposed to the solution of the quantum mechanical inverse scattering problem at fixed energy. The method relates the fixed energy phase shifts to those arising in an auxiliary Sturm-Liouville problem via the interpolation theory of the Weyl-Titchmarsh m-function. Then a Marchenko equation is solved to obtain the potential.




\end{abstract}

\pacs{02.30.Zz, 02.60.Cb, 02.60.Pn, 03.65.Nk}

\maketitle


\section{Introduction}
Inverse scattering theory based on the spectral problem of the $s-$wave Schr\"odinger equation has been worked out in the fifties \cite{Povzner,Levitan,Marchenko} and developed further in the eighties with the aim to apply to  problems in atomic, nuclear and subnuclear physics \cite{Coz,Geramb}. This type of inverse scattering theory provides the fixed-$l$ potentials derived from input spectral data (eigenvalues and normalization constants) as well as scattering phase shifts $\delta_l(k)$ known for all energy $E=(\hbar k)^2/(2m).$

Another type of inverse method has been developed in the sixties \cite{Regge,Newton,Sabatier} to recover the fixed energy (or fixed-k) potentials from a set of phase shifts  $\{\delta_{l=0,1,\ldots,l_{\text{max}}}\}$ given at a particular wave number $k$. For practical analysis of the measured scattering angular distribution, the modified Newton-Sabatier (mNS) method proved to be a powerful procedure \cite{Munchow,Eberspaecher}. Also the Cox-Thompson method, developed in the last decade to practical problems, offers results with  nice physical properties such as non-singular behavior of potential at the origin or demanding only a finite set of input data \cite{Apagyi,Palmai} to recover the potentials.

In this letter a third type of inverse method will be presented. It belongs to the fixed energy category requiring a set of phase shifts at one energy but uses spectral information belonging to an auxiliary problem resulted from a Liouville transformation. The new method bears resemblance to a recent development by Horvath and Apagyi (HA) \cite{HA,PHA}. However, whereas HA obtain results from the Gelfand-Levitan (GL) equation by solving a moment problem,  the present method makes use of the Marchenko integral equation in connection to the auxiliary spectral problem.

Because in all physical problem the potential can be known beyond a finite radius $r=a<\infty$ it is enough for a practical inverse scattering method to seek the inverse potential $q(r)$ within a finite range $0<r<a$ [beyond which $q(r>a)\equiv 0$ is assumed]. The HA method applies a variable transformation $r=a\exp{(-x)}$ and thereby casts the finite interval $[0,a]$ in  $r$ to the half line $[\infty, 0]$ in  $x$. The inverse potential is obtained from the derivative of the transformation kernel $K(x,x)$ which is determined from the GL equation within the interval $[x,0]$. This means that the potential $q(r')$ will be known in the interval $[r,a]$ corresponding to the transformed interval $[x,0]$. In order that the potential be known at the origin $r=0$, the kernel $K(x,x)$ should have been calculated accurately as $x\to\infty$. This, however, is difficult in view of the restricted number of input data.

If one tries, contrary to the HA method, to solve the inverse problem by using the Marchenko integral equation then one gets the transformation kernel $K(x,x)$ within the interval $[\infty,x]$ which suggests that a more accurate potential value can be expected at the origin $r=0$ than in the case of the HA method. As we will see this task is feasible by straightforward numerical evaluation of certain integrals. The basis of the theory remains the same as in the HA method, i.e. the recognition of the fact that the input set of fixed energy phase shifts $\{\delta_{l=0,1,\ldots,l_{\text{max}}}\}$ is related with the m-function of the auxiliary (Liouville-transformed) inverse spectral problem. While the HA method leads to the solution of a moment problem where the moments are calculated also from the input phase shifts, the present procedure requires the Jost functions to be determined for the auxiliary problem. Restricting ourselves to the simple case when there exists no bound state in the auxiliary spectral problem, we present some illuminating examples  and point out the necessary future development related to the generic case.

In section 2 the formalism will be outlined.
Here an interesting (and hitherto seemingly overlooked) connection between the GL and Marchenko equations is also discussed together with an adopted approximation of the m-function.
Section 3 contains the illustrative examples, in a part where analytic solutions are known for the auxiliary (transformed) problem and, in another part where there is no such a pre-known information available. Nevertheless, in all the cases we present also the results which are obtained without the use of the analytic solutions. These examples can help us to assess the accuracy of the underlying numerical procedure and the approximations used. Finally, section 4 is devoted to the discussion and conclusion.

\section{Theory}

Consider the radial Schr\"odinger equation (in units of $\hbar^2/2m=1$)
\begin{equation}\label{radSch}
r^2\left[-\frac{d ^2}{d  r^2}+q(r)-k^2\right]\varphi_l(r)=-l(l+1)\varphi_l(r),\quad r\in(0,a),
\end{equation}
which, after a Liouville-transformation
\begin{equation}\label{Ltransf}
y(x,-(l+1/2)^2)=r^{-1/2}\varphi_l(r),\qquad r=ae^{-x},
\end{equation}
takes the form of a standard  Sturm-Lioville (or $s-$wave Schr\"odinger) eigenvalue equation
\begin{equation}\label{StLio}
\left[-\frac{d ^2}{d x^2}+Q(x)\right]y(x,\lambda)=\lambda y(x,\lambda)\qquad x\in(0,\infty)
\end{equation}
with the transformed potential
\begin{equation}
Q(x)=r^2(q(r)-k^2),\qquad r=ae^{-x}.
\end{equation}
The m-function of the Sturm-Liouville (SL) operator $$\left[-\frac{d^2}{dx^2}+Q(x)\right]$$ is known at the point values $\lambda=-(l+1/2)^2,$ $l=0,1,2,\ldots,$ $l_{\text{max}}$ in terms of the phase shifts $\delta_l$ generated by the potential $q(r)$ [with $q(r>a)=0$] at a fixed wave number $k$, because the following relation \cite{HA} holds:
\begin{align}\label{mfunct}
 m\left(-(l+1/2)^2\right)&=\frac{y'(0,-(l+1/2)^2)}{y(0,-(l+1/2)^2)}\nonumber\\
&=-ka\frac{J'_{l+1/2}(ka)-\tan\delta_l
Y'_{l+1/2}(ka)}{J_{l+1/2}(ka)-\tan\delta_l Y_{l+1/2}(ka)}.
\end{align}

The Marchenko equation reads as
\begin{equation}\label{March}
\mathcal{F}(x+y)+\mathcal{K}(x,y)+\int_x^\infty \mathcal{K}(x,t)\mathcal{F}(t+y)dt=0\qquad (x\leq y)
\end{equation}
for the transformation kernel $\mathcal{K}(x,y)$ from which the wave function can be obtained by the Povzner-Levitan representation as
\begin{equation}\label{Povzner}
y(x,\lambda)=\frac{1}{\sqrt{\lambda}}\sin(\sqrt{\lambda}x)+\frac{1}{\sqrt{\lambda}}\int_x^ \infty \mathcal{K}(x,y) \sin(\sqrt{\lambda}y)dy.
\end{equation}
This assumes a Dirichlet boundary condition
\begin{equation}\label{hinit}
y(0,\lambda)=0,\quad y'(0,\lambda)=1,
\end{equation}
and implies the relation
\begin{equation}
Q(x)=-2\frac{d\mathcal{K}(x,x)}{dx}
\end{equation}
 between the potential and the transformation kernel which can be obtained upon insertion of (\ref{Povzner}) into (\ref{StLio}).

The input function  $\mathcal{F}(x)$ to the Marchenko equation  (\ref{March}) contains contributions of the bound states (discrete spectrum) and scattering states (continuous spectrum)
\begin{equation}\label{FMarch}
\mathcal{F}(x)=\sum_{j=1}^B \frac{1}{m_j} e^{-\lambda_j x}+\frac{1}{2\pi}\int_{-\infty}^\infty\left(1-e^{2i\Delta(\kappa)}\right)e^{i\kappa x}d\kappa
\end{equation}
where $B$ denotes the number of the bound states supported by the potential $Q(x)$,
and $\lambda_j<0$, $ m_j>0$ denote, respectively, the eigenvalues and normalization constants of the eigenfunctions $y(x,\lambda_j)$ belonging to the Schr\"odinger equation (\ref{StLio}).
The phase function $\Delta(\kappa)$ corresponds to the ($s-$wave) phase shift
arising when a particle with wave number $\kappa$ is scattered by the hypothetical potential $Q(x)$.

It is related to the modulus of the Jost function $f^+(\kappa)=|f^+(\kappa)|e^{i\Delta(\kappa)}$ by the dispersion relation \cite{Levitanbook}
\begin{equation}\label{Delta}
\Delta(\kappa)=\frac{1}{\pi}\mathcal{P}\int_{-\infty}^\infty \frac{\log |f^+(\kappa')|d\kappa'}{\kappa'-\kappa}.
\end{equation}
The modulus, in turn, is connected to the m-function by
\begin{equation}\label{fmod-m}
\frac{|f^+(\kappa)|^2}{\kappa}=\lim_{\varepsilon\to0^+}\frac{1}{{\rm Im}\, m(\kappa^2+i\varepsilon)},\qquad \kappa>0.
\end{equation}
The latter equality can easily be proved by using the reflection principle $\overline{m(z)}=m(\overline{z})$ and the Wronskian property $f^-(f^+)'-(f^-)'f^+=2ik$ of the two linearly independent Jost functions $f^{\pm}$.

In summary, if the m-function is known near the positive half axis of its argument one can determine the potential $Q$ (or $q$).

\subsection{Connection to the Regge-Loeffel-Sabatier-Levitan framework}

Let us transform back both the input and transformation kernels into the r-space. Performing the transformation $x=-\log\frac{r}{a}$  and using the notations
\begin{align}
\tilde{\mathcal{K}}(r,r')&=\mathcal{K}\left(-\log\frac{r}{a},-\log\frac{r'}{a}\right)\\
\tilde{\mathcal{F}}(r,r')&=\mathcal{F}\left(-\log\frac{r}{a}-\log\frac{r'}{a}\right)=\mathcal{F}\left(-\log\frac{rr'}{a^2}\right)
\end{align}
we get from the Marchenko integral equation (\ref{March}) the following equation of Gelfand-Levitan type
\begin{equation}\label{M1}
\tilde{\mathcal{F}}(r,r')+\tilde{\mathcal{K}}(r,r')+\int_0^r \tilde{\mathcal{K}}(r,r'')\tilde{\mathcal{F}}(r'',r')\frac{dr''}{r''}=0\quad (r\geq r').
\end{equation}
The (wanted) fixed energy potential can also  be calculated as
\begin{equation}\label{potM}
q(r)=\frac{2}{r}\frac{d\tilde{\mathcal{K}}(r,r)}{dr}+k^2.
\end{equation}
Note that equations (\ref{M1}) and (\ref{potM}) are exactly the same as those appearing in the Regge-Loeffel-Sabatier-Levitan framework of inverse scattering theory at fixed energy (see Chapter 5 of \cite{Levitanbook} where $\tilde{\mathcal{F}}(r,r')=F_{\Delta_0,D}(r,r')$ with $k=1$). The difference is that we have here a new method to calculate the exact input kernel in terms of the fixed energy phase shifts through the m-function of the auxiliary Schr\"odinger  operator.

Note that equation (\ref{M1}) can be transformed to the Gelfand-Levitan form by a further transformation of $\tilde{\mathcal{F}}$ and $\tilde{\mathcal{K}}$.

\subsection{Interpolation of the m-function}
 As relation (\ref{fmod-m}) shows, the m-function must be evaluated at argument values which are different from those offered by the input data as showed by equation (\ref{mfunct}). Therefore, we must use interpolation theories for calculating the m-function. These interpolation theories all rely upon
 a consequence of Simon's representation
of the m-function \cite{Simon}. It states that the m-function can be represented essentially by a
Laplace transform. For both the Fourier and the Laplace transforms there are known
interpolation formulas, and the latter, a recent work of Rybkin and Tuan \cite{tuan}, will be adopted for our purposes.

Thus, we shall use the following interpolation formula for the m-function (special case of Theorem 5 of Ref. \cite{tuan}):
\begin{equation}\label{minter}
m(\lambda)\approx i\sqrt{\lambda}+\sum_{n=0}^{l_{\text{max}}} c_n(-i\sqrt{\lambda})\sum_{m=0}^n a_{nm}(m(-\omega_m^2)+\omega_m)
\end{equation}
with $\omega_m=m+\frac{1}{2}$ and
\begin{equation}
c_n(x)=(2n+1)\frac{\left(\frac{1}{2}-x\right)_n}{\left(\frac{1}{2}+x\right)_n},\quad
a_{nm}=\frac{(-n)_m(n+1)_m}{(m!)^2}.
\end{equation}
$(x)_n$ is the Pochhammer symbol:
\begin{equation}
(x)_n=x(x+1)\ldots(x+n-1),\qquad (x)_0=1.
\end{equation}

The interpolation formula applies for locally absolute integrable potentials. Its domain of convergence depends on the specific features of the potential and we suppose it to be valid on the whole complex $\lambda-$plane throughout the subsequent numerical examples.

\section{Applications}
As an illustration of the above theory we show first examples where solutions of the transformed problem are known analytically. Such are the cases of the constant potentials and the step potentials. Then we exhibit some results for cases where the solution of the transformed problem is not at hand, and thus the interpolation formula (\ref{minter}) must be applied. The selected example for the latter case is the shifted truncated Coulomb potential problem, whose reconstruction is known to be a challenging exercise for any quantum inverse scattering method \cite{Hilk}.

\subsection{Constant potentials}

Let us reconstruct first the constant potential
\begin{equation}\label{constpot}
q(r)=\begin{cases} q_0& \mbox{for }r\leq a \\ 0& \mbox{for } r>a\end{cases}
\end{equation}
which generates the transformed potential
\begin{equation}
Q(x)=-se^{-2 x},\qquad s=a^2(k^2-q_0).
\end{equation}
The solution of the transformed Schr\"odinger equation (\ref{StLio}) becomes
\begin{equation}\label{constsol}
\psi(x)=C_1 J_{i\sqrt{\lambda}}\left(\sqrt{s}e ^{-x}\right)+C_2
J_{-i\sqrt{\lambda}}\left(\sqrt{s}e ^{-x}\right)
\end{equation}
from which the m-function being the logarithmic derivative of the unique $L^2$ solution at the origin takes the form
\begin{equation}\label{m-const}
m(\lambda)=\frac{\psi'(0)}{\psi(0)}=-\sqrt{s}\frac{J_{-i\sqrt{\lambda}
}'(\sqrt{s})}{J_{-i\sqrt{\lambda}}(\sqrt{s})}.
\end{equation}
The ($s-$wave) phase shift belonging to equation (\ref{StLio}) can be expressed from (\ref{constsol}) as ($\kappa=\sqrt{\lambda}$)
\begin{equation}\label{Deltaconst}
\Delta(\kappa)=i \tanh ^{-1}\left(\frac{1-\frac{4^{i \kappa} s^{-i
   \kappa} \Gamma \left(i \kappa+1\right) J_{i
   \kappa}\left(\sqrt{s}\right)}{\Gamma \left(1-i
   \kappa\right) J_{-i \kappa}\left(\sqrt{s}\right)}}{1+\frac{4^{i \sqrt{\lambda }} s^{-i
  \kappa} \Gamma \left(i \kappa+1\right) J_{i
  \kappa}\left(\sqrt{s}\right)}{\Gamma \left(1-i
  \kappa\right) J_{-i \kappa}\left(\sqrt{s}\right)}}\right).
\end{equation}

\begin{figure}
\centering
\subfigure{\includegraphics[scale=0.48]{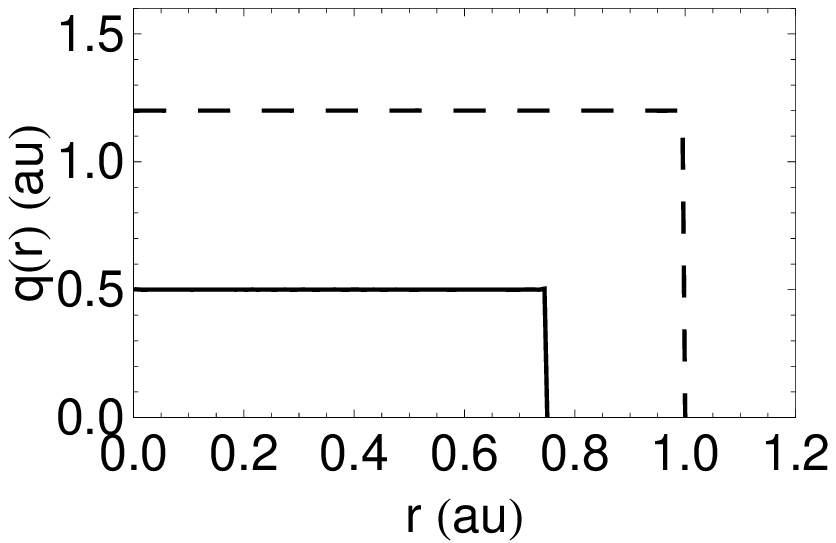}}
\subfigure{\includegraphics[scale=0.48]{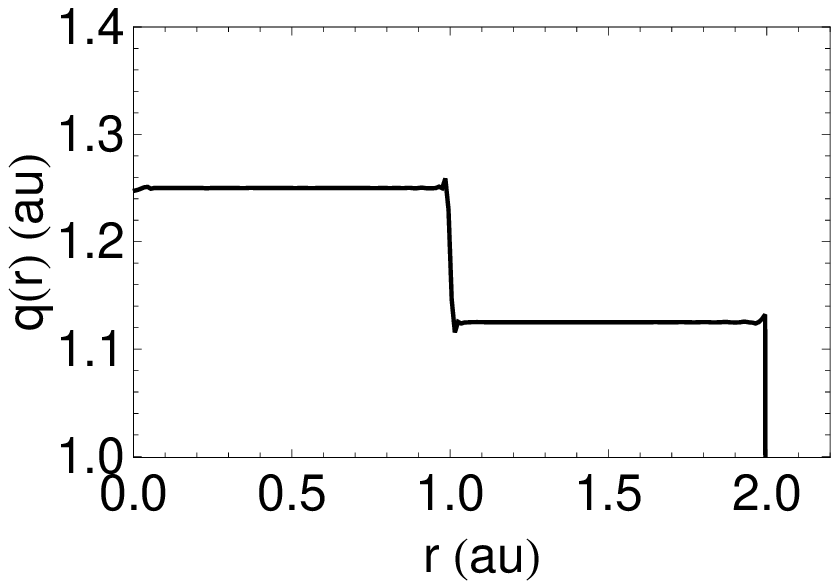}}\\
\subfigure{\includegraphics[scale=0.48]{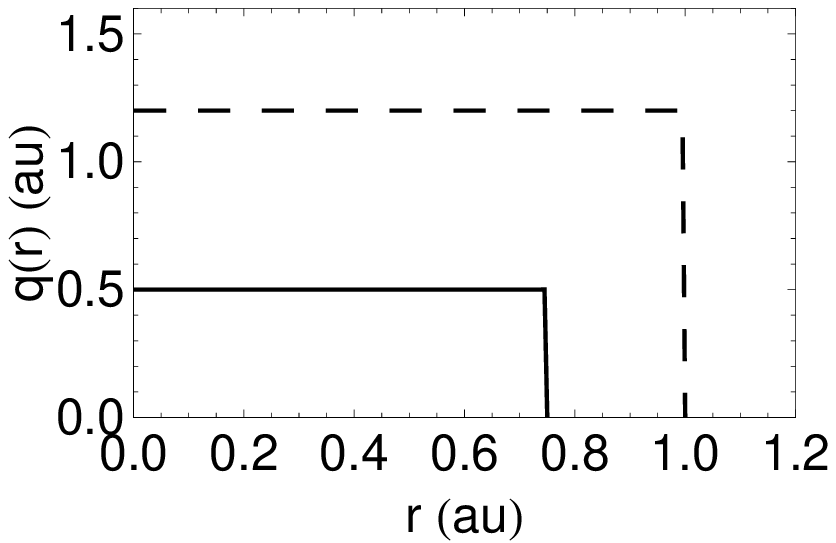}}
\subfigure{\includegraphics[scale=0.48]{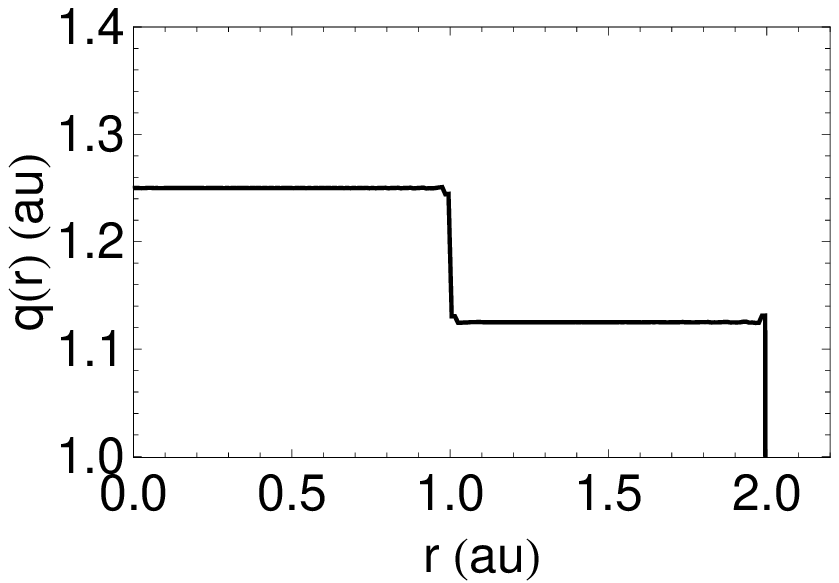}}\\
\subfigure{\includegraphics[scale=0.48]{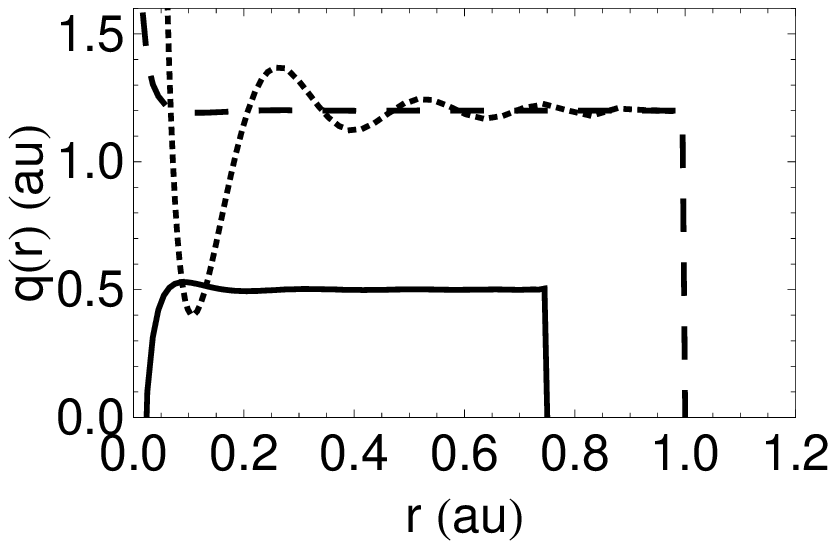}}
\subfigure{\includegraphics[scale=0.48]{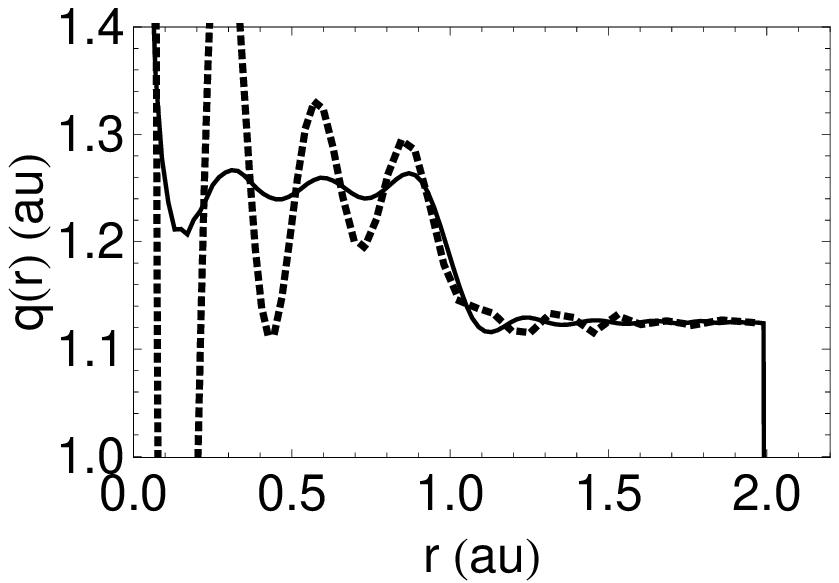}}
\caption{Recovery of the constant (\ref{constpot}) with parameters $q_0=1.2$,
 $a=1$ (dashed lines) and $q_0=0.5$, $a=0.75$ (continuous lines) and step (\ref{steppot}) potentials using exact input m-function (upper row), exact phase shift functions (middle row) and fixed energy phase shifts (lower row). The HA results are shown as dotted line. (For further details see text.)}\label{fig1}
\end{figure}

Now, we are in the position to perform the calculation at three different levels. Either we can use
 the exact m-function (\ref{m-const})
 or
 the exact phase function (\ref{Deltaconst})
  or
  the interpolation formula (\ref{minter}) for an approximate calculation of the m-function using the set of input phase shifts $\{\delta_l\}$.
Results of the three performances are depicted in Figure \ref{fig1} (left panel) for two constant potentials with $q_0=1.2$ and  $a=1$ (dashed lines), and $q_0=0.5$ and $a=0.75$ (continuous lines). The wave number in each case is chosen to be $k=1$. Eleven  input phase shifts ($l_{\text{max}}=10$) are used when the interpolation formula is applied (see Fig. \ref{fig1}). Apparently, while there is no practical difference between the calculations performed at the levels with exact m-function and $\Delta$-function, the weakest point appears to be the use of the interpolation formula. As a comparison we have included in Fig. \ref{fig1} as dotted line the result given by the HA method \cite{HA}.

%

\subsection{Step potential}
As a next example we reconstruct the step potential
\begin{equation}\label{steppot}
q(r)=
\begin{cases}q_1 &\mbox{for } r\leq r_0\\
q_2 & \mbox{for } r_0\leq r\leq a\\
0 & \mbox{for } r>a,\end{cases}
\end{equation}
with parameters $q_1=1.25, q_2=1.125, r_0=1,a=2$.
The transformed potential looks like
\begin{equation}
Q(x)=\begin{cases} -s_2e^{-2x} &\mbox{for } x\leq x_0\\
-s_1e^{-2x} & \mbox{for } x> x_0, \end{cases}
\end{equation}
and the transformed Schr\"odinger equation (\ref{StLio}) admits the solution
\begin{equation}
\psi(x)=\begin{cases}
C_1 J_{i\sqrt{\lambda}}\left(\sqrt{s_2}e ^{-x}\right)+C_2 J_{-i\sqrt{\lambda}}\left(\sqrt{s_2}e ^{-x}\right), \,x\leq x_0,\\
D_1 J_{i\sqrt{\lambda}}\left(\sqrt{s_1}e ^{-x}\right)+D_2 J_{-i\sqrt{\lambda}}\left(\sqrt{s_1}e ^{-x}\right), \,x\geq x_0.
\end{cases}
\end{equation}
From this the m-function takes the form
\begin{equation}
m(\lambda)=\frac{J_{i\sqrt{\lambda}}'(\sqrt{s_1})
W[1_{-},2_{-}]
-J_{-i\sqrt{\lambda}}'(\sqrt{s_1})
W[1_+,2_-]}
{J_{i\sqrt{\lambda}}(\sqrt{s_1})
W[1_-,2_-]
-J_{-i\sqrt{\lambda}}(\sqrt{s_1})
W[1_{+},2_{-}]}
\end{equation}
where
\begin{equation}
W[1_\pm,2_\pm]=W[J_{\pm i\sqrt{\lambda}}^{s_1},J_{\pm i\sqrt{\lambda}}^{s_2}](e^{x_0}), \quad J_\nu^\alpha(x)\equiv J_\nu(\sqrt{\alpha}x)
\end{equation}
with the Wronskian $W[f,g](x)=f(x)g'(x)-f'(x)g(x)$. The $s$-wave phase function can be written as ($\kappa=\sqrt{\lambda}$)
\begin{equation}
\Delta(\kappa)=i \tanh ^{-1}\left(\frac{1+\frac{4^{i \kappa} s_1^{-i \kappa} \Gamma \left(i \kappa+1\right)}{\Gamma
   \left(1-i \kappa\right)}H}{1-\frac{4^{i\kappa} s_1^{-i \kappa} \Gamma \left(i \kappa+1\right)}{\Gamma \left(1-i \kappa\right)}H}\right)
\end{equation}
with
\begin{equation}
H=\frac{J_{-i\kappa}(\sqrt{s_2})
W[1_+,2_+]-J_{i\kappa}(\sqrt{s_2})
W[1_+,2_-]}
{J_{i\kappa}(\sqrt{s_2})
W[1_-,2_-]-J_{-i\kappa}(\sqrt{s_2})
W[1_-,2_+]}.
\end{equation}

The results obtained at $k=1$ are depicted in Figure \ref{fig1} (right panel) at various levels of approximation indicated. Also in this case, the less satisfactory result is obtained by use of the interpolation formula (with $l_{\text{max}}=20$) and, as comparison, the result of HA method is shown by dotted line.

\subsection{Shifted truncated Coulomb potential}

We now reconstruct the shifted truncated Coulomb potential defined by
\begin{equation}\label{shiftedcoul}
q(r)=\begin{cases}
\frac{A}{r}-\frac{A}{a}&\mbox{for } r\leq a\\
0& \mbox{for } r>a
\end{cases}
\end{equation}
with the parameters $a=2$ and $A=1$. Requiring a continuous logarithmic derivative of the associated wave functions leads to the formula for the input phase shifts
\begin{align}\label{phatra}
\delta_l&=\tan ^{-1}\left(\frac{u_l'(ka)-C_lu_l(ka)}{v_l'(ka)-C_lv_l(ka)}\right),\\
C_l&=\frac{k_BF'_l(k_Ba,\eta)}{k F_l(k_Ba,\eta)},
\end{align}
with $k_B^2=k^2+A/a,\eta=A/(2k_B)$, and the usual $u_l, v_l$ Riccati-Bessel and $F_l$ Coulomb wave functions.

\begin{figure}
\centering
\subfigure[$l_{\text{max}}=1, k=0.8$]{
\includegraphics[scale=0.47]{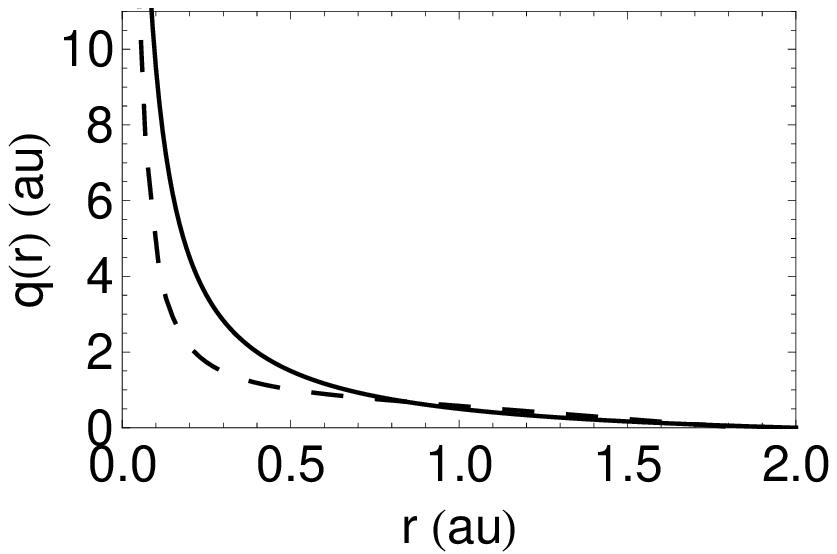}
}
\subfigure[$l_{\text{max}}=3, k=1.0$]{
\includegraphics[scale=0.47]{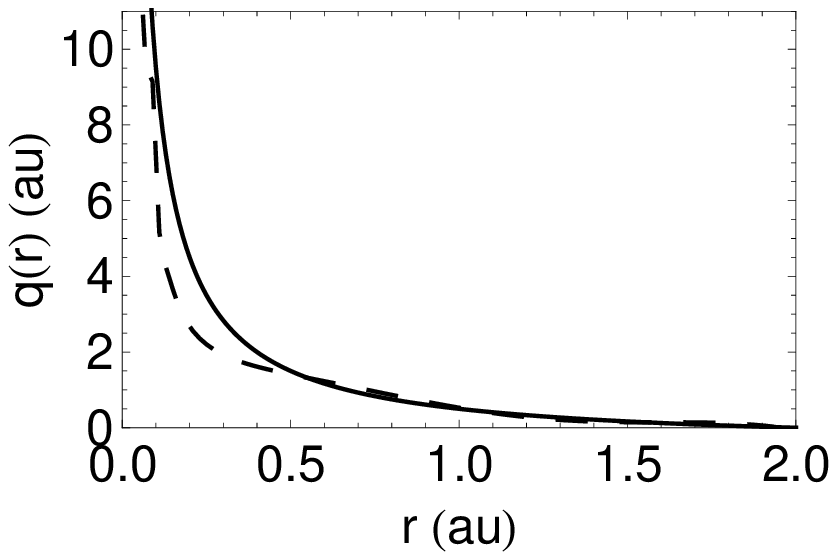}
}
\caption{Recovery of shifted truncated Coulomb potential (\ref{shiftedcoul}) at two different wave numbers and sets of input data indicated.}\label{fig2}
\end{figure}

Because in this case there is no analytic result to the m-function or phase function $\Delta$ of the transformed problem (\ref{StLio}), we have the only possibility to apply the Rybkin-Tuan formula (\ref{minter}) for calculating the m-function.
The calculation  has been performed  at two different settings of the wave number $k=0.8$ and $k=1.0$.
The corresponding input phase shifts calculated from equation (\ref{phatra}) are listed in Table 1.
The results exhibited in Figure \ref{fig2} show an improvement of the procedure with increasing energy as more input phase shifts are involved, according to the semiclassical thumb rule $l_{\text{max}}\approx ka $. However, the reproduction shows, in all cases, a generic departure from the exact result which is attributed to the shortcomings with use of the interpolation formula (\ref{minter}).

\begin{table}[h!]
\begin{center}
\begin{tabular}{rrrr}
\hline
\multicolumn{2}{c}{$k=0.8$} & \multicolumn{2}{c}{$k=1.0$}\\
\cline{1-2}\cline{3-4}
$ l$ & $\delta_ l$&$ l$ & $\delta_ l$\\
\hline
$0$ &$-0.2991$&$0$ &$-0.3481$\\
$1$ &$-0.0317$&$1$ &$-0.0538$\\
&                        &$2$ &$-0.0046$\\
&                        &$3$ &$-0.0002$\\
\hline
\end{tabular}
\end{center}
\caption{Input phase shifts $\delta_l$ calculated from (\ref{phatra}) for the shifted truncated  Coulomb potential}
\label{coulomb}
\end{table}

\section{Discussion and conclusion}

Using the Marchenko integral equation (\ref{March}) a new fixed energy inverse quantum scattering method is presented for the recovery of potentials bounded to a finite range $r\in [0,a]$, beyond which it is negligible. The method consists of Liouville transforming the radial variable and the radial wave function (see eq. (\ref{Ltransf})), and thus casting the problem of inverse scattering at fixed energy to the inverse spectral problem of a Sturm-Liouville operator on the half line $x\in[\infty,0]$. The possibility of doing this is based on the observation \cite{HA} that the set of input phase shifts can be related to the m-function of the transformed problem. In \cite{HA} the Gelfand-Levitan equation has been used to the solution of the transformed problem, from which one recovers the potential in the interval  $\lim_{ r\to 0}[r,a]$. However one is frequently interested in the behavior of the potential at the origin $r=0$, therefore, the usage of the Marchenko equation is proposed  in this work, because it recovers the potential (after back-transforming it) in the interval $\lim_{r\to a}[0,r]$. Thus, it has been assumed, in view of the finite number of input data, that the present method recovers the  potential  at the origin more accurately than that proposed in Ref.\cite{HA}.

By studying simple examples we have shown that it is, in fact, so. In the cases where we know the exact m-function (or phase-function), the present procedure is capable to recover the potential at the origin exactly while the method  of Ref. \cite{HA} is not.

The latter method has been formulated \cite{HA} to accommodate to usage of input phase shifts by solving a moment problem; there is no room (yet, in the present stage of the formalism) to offer any entry points for usage of exact (intermediate) input data, such as, the m-function and/or the $\Delta$ phase function.
The moment problem can only be solved approximately even if these intermediate quantities were at hand. This feature and the requirement of knowledge of the transformation kernel on the (infinite) half line leads the HA method always to produce a worse (re)construction of  the potential at the origin.

The present method is conducted on a way paved by analytical formulas derived from standard scattering and spectral theory. It offers explicit entry points for applying exact intermediate input data (if known).
However, in proportion to the moment problem of Ref.\cite{HA}, the present technique also needs to use approximations in realistic situations. The determination of the m-function from the input data phase shifts $\{\delta_l\}$ [related to the wanted fixed energy potential $q(r)$] can be done through the application of the interpolation theory of the m-function, an up to date topic of the mathematical physics today.
We have adopted the theory worked out by Rybkin and Tuan \cite{tuan}
which is an important development based on an earlier study \cite{tuan1}.
Closer inspection of their formula (\ref{minter}) reveals that, strictly speaking, it might not be applicable to our purpose: the choice $\omega_m=m+1/2$ in their framework is only permitted for cases when the norm of the potential $Q$ is small enough and even then the domain of convergence of (\ref{minter}) is such that it excludes the vicinity of the positive real half-line. However these restrictions originate from bounds on the domain of convergence of Simon's formula (overestimating the potential by a constant) and are expected to be relaxed when the exponential decay of $Q$ is taken into account. The fact that our numerical examples recover reasonable results (see Figs. \ref{fig1} and \ref{fig2}) indicates this. At the same time we recognize that an important development of the present inverse method relies upon the more thorough examination of Simon's representation and the interpolation theory of the m-function.

Another important development of the present method would be the inclusion of the intermediate bound states into the procedure. This could be done within the above formalism but we left this investigation to a future work.



\end{document}